# ATCA LLRF SYSTEM FOR ASU COMPACT FEL *

A. Young, J. M. D'Ewart, J. Frisch, B. Hong, T. Straumann, M. Weaver, C. Xu

SLAC National Accelerator Laboratory, Menlo Park, USA

*Abstract*

SLAC National Accelerator Laboratory is collaborating with Arizona State University to design a LLRF system towards the advancement of the ASU Compact X-ray Light Source (CXLS). The CXLS is a phased project to develop small brilliant x-ray sources that may produce coherent x-rays appropriate as a seed source for LCLS and LCLS II. The CXLS electron beam is produced by an x-band photoinjector and linac operating at 9.3 GHz RF frequency and 1 kHz repetition rate with up to 300 pC per bunch. The linac consists of 3 short 20-cell sections of standing wave linac. The RF gun and linac sections are powered by two 6 MW transmitters using L3 klystrons and Scandinova K1 modulators. The accelerator will initially operate in single-bunch mode and may be upgraded to multi-bunch operation at a later date. SLAC National Laboratory has developed a LLRF system in Advanced Telecommunication Computing Architecture (ATCA) for the ASU project. This talk will discuss the design of the LLRF for this project and discuss the ATCA implementation.

## OVERVIEW

The CXLS is being built at Arizona State University as part of the Biodesign Institute. The X-rays will generated by inverse Compton scattering with collisions of 200mJ, 1ps, infrared laser pulse with 300fs electron bunches at a maximum energy of 35Mev at a maximum repetition rate of 1kHz. The Accelerator system block is shown below in Fig. 1. The first component is a UV photocathode laser followed by a 4.5 cell photo injector [1] accelerating the beam to 4Mev. Next comes the first accelerating structure L1 which is a 0.35m 20 cell standing wave structure [2]. This structure is powered by the first klystron. The Klystron operated at 9.3GHz and produces a 700ns. This structure accelerates the beam to 4-12Mev. The next two structures (L2 and L3) with accelerate the beam to the final energy of 35MeV. A bunch compression chicane an also be used for bunch compression to produce electron bunches less than 100 fs. Following the chicane is the IR interaction point and an X-Ray shaping device for the experimental area.

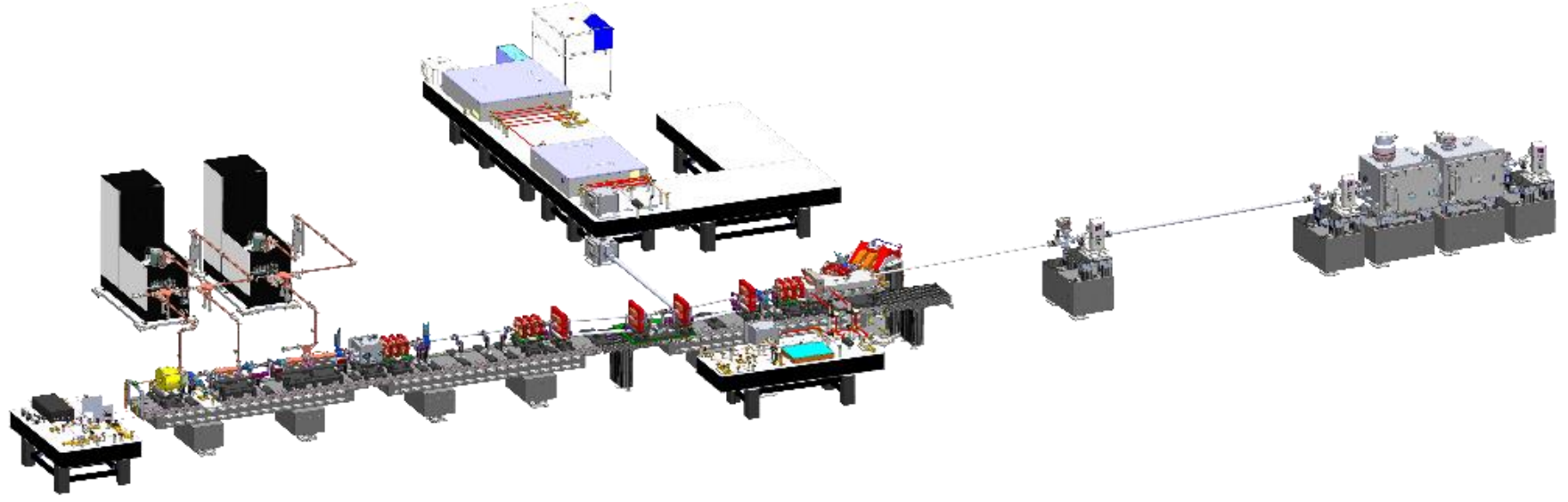

Figure 1: ASU Compact X-Ray Light Source

.

The LLRF system originally was designed at SLAC to mitigate these effects and to provide for upgradeability well into the future, the US Department of Energy is investing in upgrading the system using modern FPGA based technology. During the design process, many platforms were evaluated for applicability and maintainability. We evaluated: VME, PXI, mTCA for physics, ATCA, and conventional chassis. At the same time this evaluation was going on for LLRF upgrades at SLAC, the diagnostics, machine protection and other groups were evaluating their own common platform solution and had decided on ATCA as their platform of choice. Because of that decision and our own evaluations, it was decided that the system will be designed into the ATCA platform. SLAC and ASU have formed a collaborative partnership through a strategic partnership project (SPP). SLAC with design the LLRF system and ASU will write the software to operate the system. SLAC will design a timing system, master oscillator system, and a LLRF system. The Master Oscillator system is illustrated in figure 2 and shows how all the frequencies are derived from the laser frequency of 2325MHz. A high level block diagram of the new LLRF system is shown in Fig. 3.

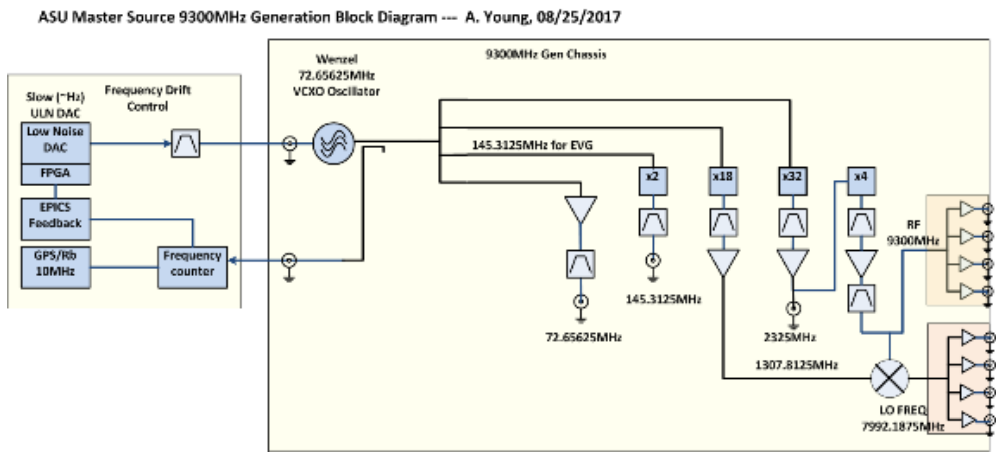

Figure 2. Master Oscillator

* Work supported by DOE contract DE-AC02-76SF00515 and SPP18-0002-SP email address: ayoung@slac.stanford.edu

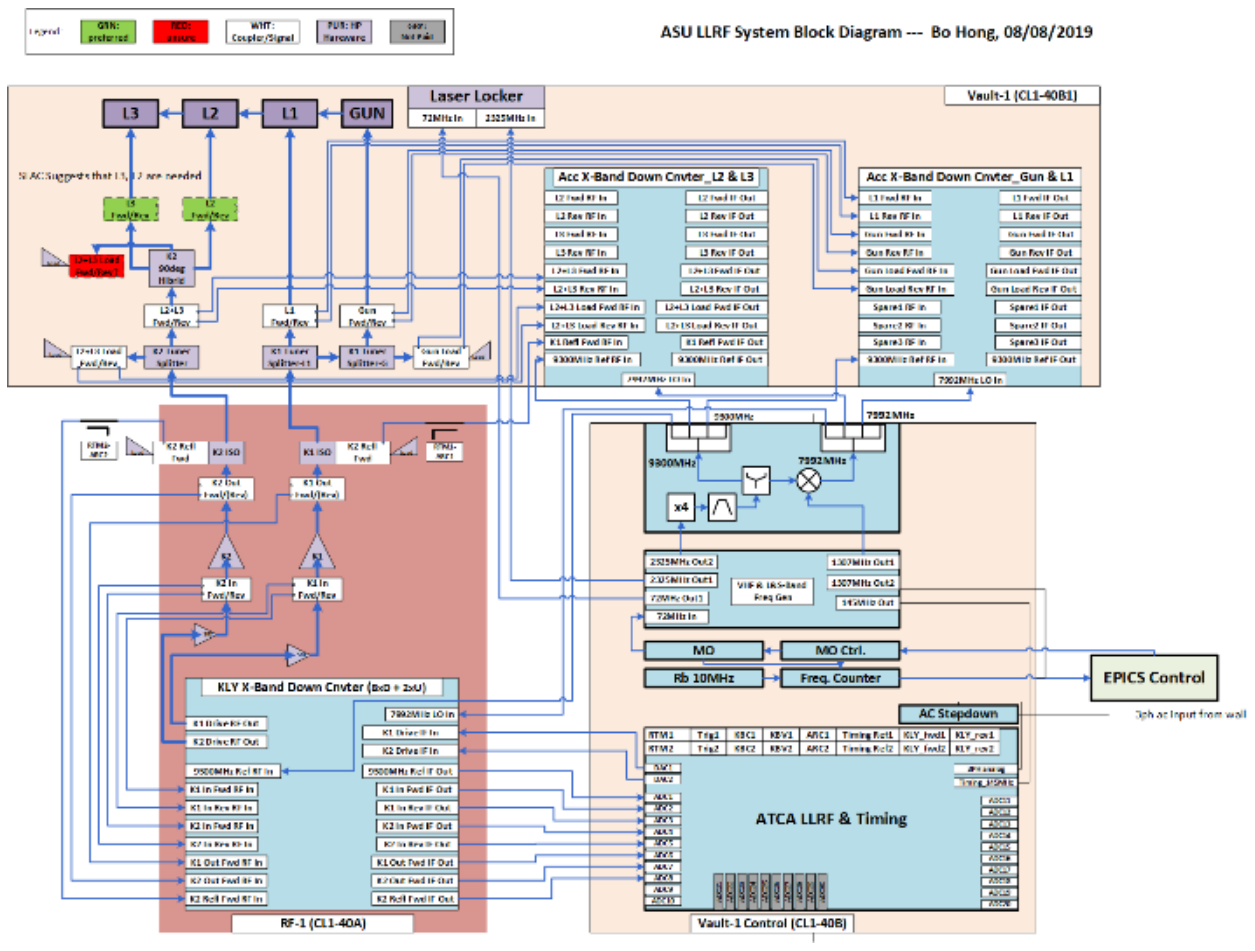

Figure 2: LLRF System at ASU.

## DETAILS OF NEW SYSTEM

The system is being designed into a new platform based upon the telecommunications system ATCA [3]. The system incorporates two advanced mezzanine cards (AMC cards) which plug into a carrier card that contains the FPGA for all signal processing functions. The overall common platform system allows for enough flexibility that the design can be used for other diagnostic and machine protection functions simply by designing other AMC cards with the specific functionality required for those systems.

The important specifications of the LLRF system are shown in table 1.

Table 1: System Specifications

| SPEC | VALUE |
|---|---|
| Noise (phase) | < 0.01 degrees 1 MHz BW |
| Noise (amplitude) | < 0.01% 1 MHz BW |
| Drift (phase) | 0.1 degrees |
| Drift (amplitude) | 0.1% (1min) (2 degrees C) |
| RF Channel Bandwidth | >10 MHz |
| RF Channel Resolution | 16 bit resolution |
| Non-linearity | 0.1 degrees for 6 dB change |
| Modulator Voltage Readback | 14 bits resolution |

### *SLAC ATCA Common Platform*

The SLAC ATCA common platform has been developed for new controls and data acquisition projects at SLAC. LCLS II and SLAC mission readiness project are based on the common platform hardware. A standard installation consists of an ATCA crate and number of application AMC carriers. An industrial CPU is used as an IOC and communicates with each carrier via 10 gigabit Ethernet.

### *Detailed Hardware Description*

In order to provide enough channels for all potential uses, we determined that 10 channels of RF down-conversion, 1 Channel of RF up-conversion and 2 Channels of Baseband signal detection were necessary. It was further determined that the interlock functions would be designed in a new ATCA rear transition module (RTM) and the klystron modulator control would be handled by ScandiNova, model K100. A more detailed block diagram of the final implementation is shown in Fig. 3.

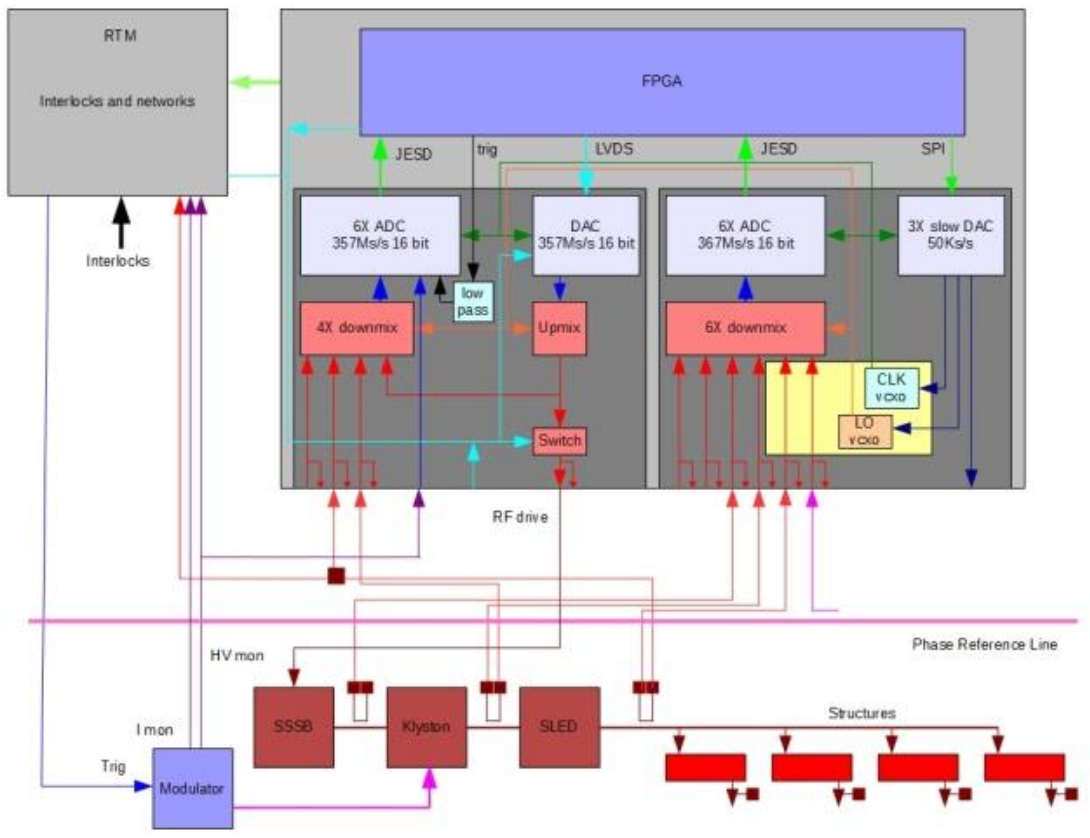

Figure 3: Detailed Block Diagram

To enable handling of the multiple clocks and LO's necessary to down-convert all SLAC frequencies it was decided that a daughter card piggy-backed on one of the AMC cards would be used to provide clock and LO signals to both AMC cards. A new pulsed solid-state sub-booster amplifier will be designed to drive each of the 8 klystrons per sector.

**AMC Cards** The two AMC cards will use the same basic architecture throughout. Each card will contain 3 ADC chips running at approximately 363 Msamp/sec. Each chip contains 2 ADCs for a total of 6 ADC channels per board. Because there simply was not enough pins to run parallel bit ADCs we chose to use the relatively new standard of JESD204B serial lanes.

We further broke the design down into what we called a precision card and an up-converter card. The Precision card will be used for comparing critical feedback signals to a reference signal from a phase reference line as shown in Fig 2. This is similar to what the LLRF team designing the LCLS-II system is doing, however we are doing it for differing reasons. Our main reason for splitting the cards in this manner is to get the maximum dynamic range achievable without interference from the signal generation card. Any other precision signals should be input to this card as well. The up-converter card will be used to create the corrected (or feedback) signal that will be fed to the structures (or cavity) depending on application.

The precision card contains six identical RF down-converters with a block diagram as shown in Fig. 4.

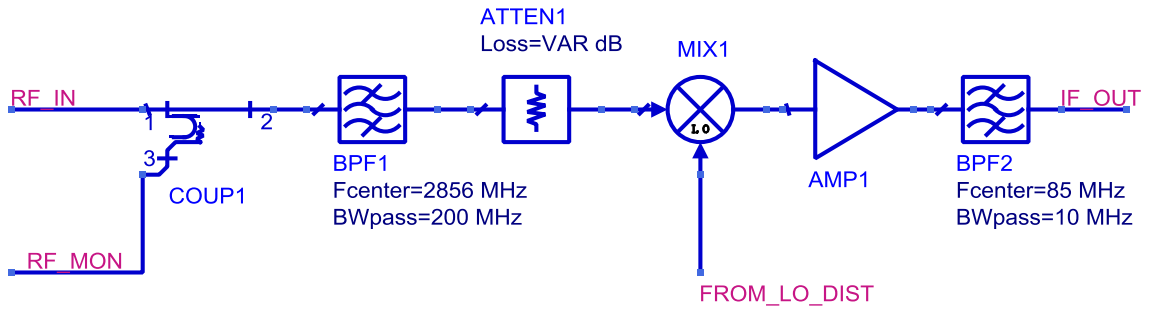

Figure 4: RF Down-converter Block Diagram

The up-converter card contains 4 down-converters identical to the ones used on the precision card as well as two buffered baseband signals. One of the baseband signals will be used to monitor the klystron modulator voltage and the other is used internally for timing synchronization of the RF pulse amplitude. The up-converter card also contains a 16 Bit parallel DAC. We had to use a parallel DAC instead of a JESD204B DAC because during initial testing we found excessive latency with the original DAC we had planned on using. The DAC outputs an IF at approximately 86 MHz and uses the same LO as the down-converters to up-convert to 1307 MHz.

**Interlock RTM** The interlock rear transition module's, (RTM) main function is to protect the klystron from various conditions that could result in destruction of the klystron. The new interlock RTM will handle the fast interlocks while the slow interlocks (water, vacuum, Magnet, temperature etc.) will be handled in the modulator itself. Fig. 5 shows a block diagram of the interlock scheme.

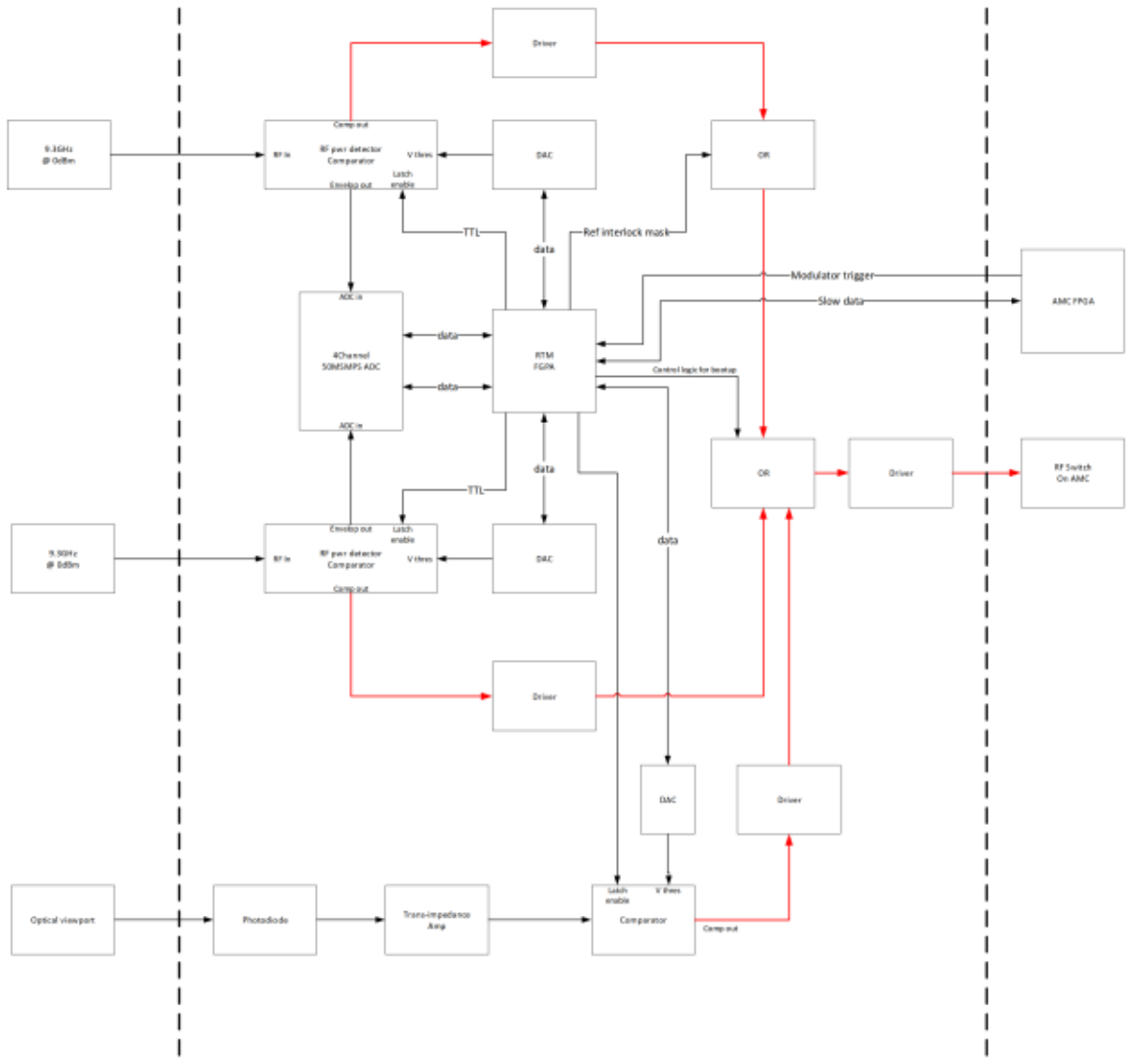

Figure 5: Interlock Scheme

## *Detailed Firmware Description*

The firmware for the LLRF scheme for the SLAC LINAC is a very complex design with many facets (Timing, amplitude and phase adjustments). Over the many years of SLAC operation, many feedbacks have been added on top of the original system to allow for much more stable operation than the original SLAC linac and as a result, the new system must fit seamlessly into those feedbacks with no adverse effects.

All of the main controllability of the system relies on individual klystron phase and amplitude control. To achieve this, a fiducial riding on top of the 476 MHz was used to time the pulse with the arrival of the beam. In that way there was a guarantee that the trigger would be timed in phase with the RF signal being amplified and distributed to the structures of the main linac. With the addition of LCLS-I to SLAC, a new even system was added [4] which essentially split the timing into two differing systems; One being the fiducial system and one being the event system. The event system is used to time a myriad of other systems around the machine (BPM's cameras etc) One goal of this new system is to combine the timing into one system only with a fiber going to each station with the timing information and as a result complicates the firmware implementation.

To achieve pulse to pulse timing stability of less than 100ps on the envelop of the RF, we need a way to detect the trigger at much less than the roughly 2.7ns time allowed by our system clock. To achieve sub-clock sample timing, a clever interpolation scheme was created by running the trigger through a low pass filter, sending to an ADC and then using DSP interpolation to detect the zero crossing with sub time sample resolution. This new timing information is then used to create an interpolated baseband signal which is multiplied (digitally) onto a digitally controlled oscillator which is phase locked to the incoming phase reference IF. A block diagram of the overall firmware scheme is shown in Fig 6.

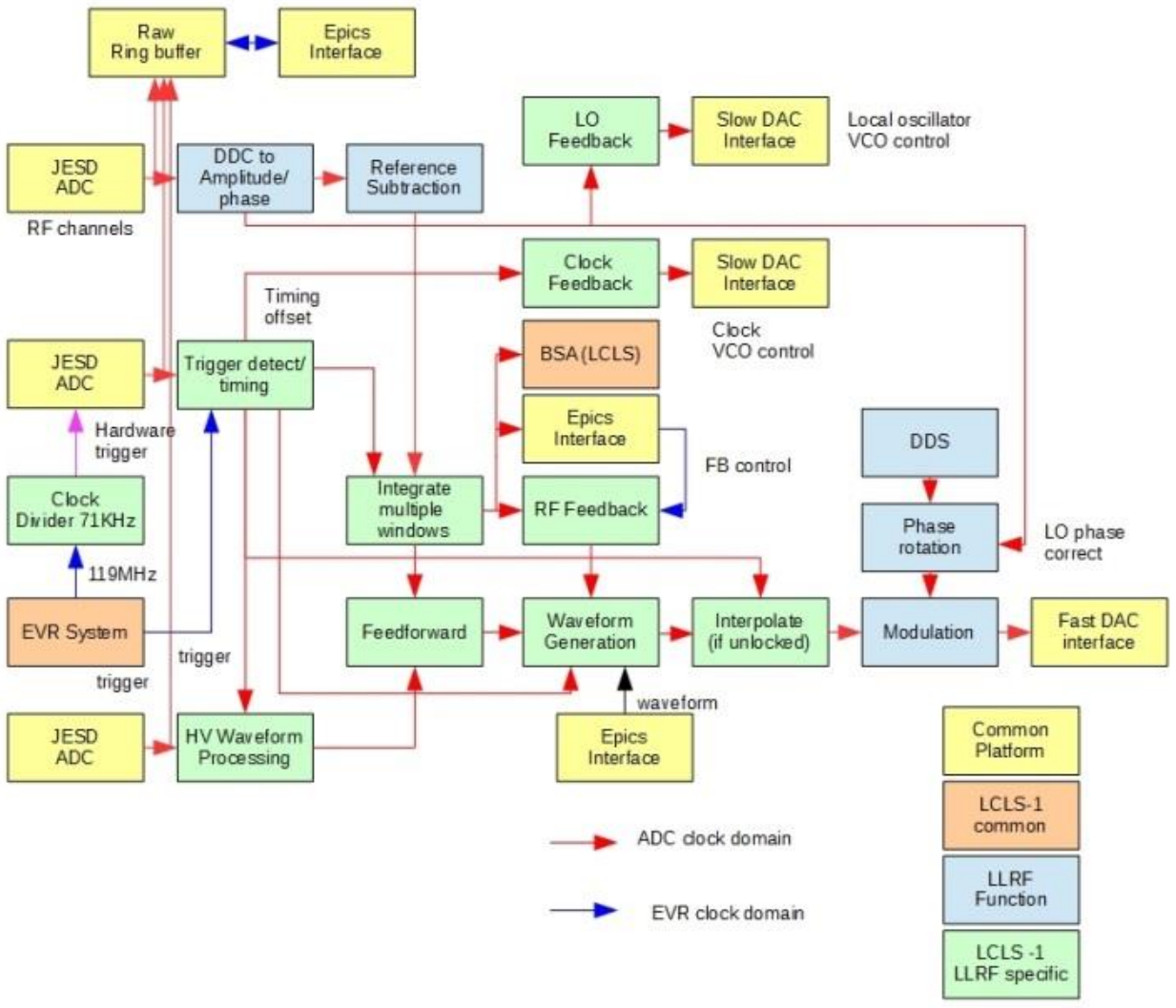

Figure 6: Detailed firmware Block diagram.

# PROTOTYPE RESULTS

Prototype cards for the ASU LLRF system have been built and initial firmware has been developed to allow for data taking and offline data processing. We present some of the results here.

**Phase Accuracy** To measure phase accuracy, a common signal is sent into two channels of down-conversion, sampled at an IF/Sample ratio of 5/21 (for near IQ sampling [x5]. The near IQ down conversion is filtered by a simple 21 average filter and then digitally down converted to baseband. The LO is locked to the IF of the incoming 1307 MHz signal and the clock is locked using one ADC channel which is sampling an external 1307 MHz signal. The locking and measurement scheme is shown below in Fig. 5.

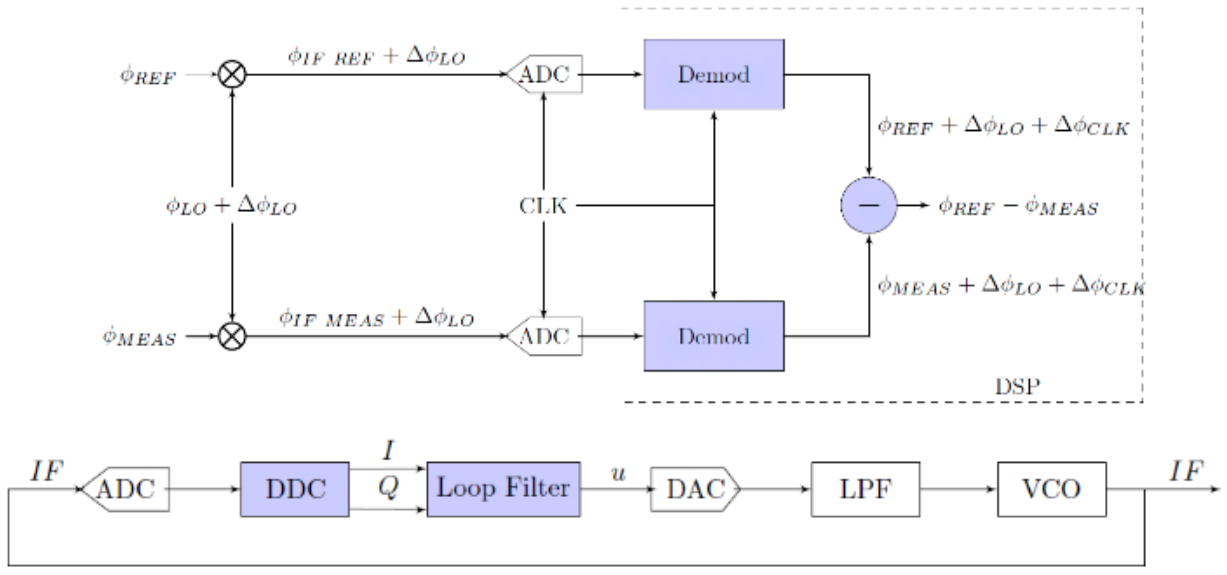

Figure 5: Phase measurement and locking scheme.

To calculate the base phase error, the two IQ signals from ADC2 and ADC3 are converted to phase and amplitude using a cordic algorithm in matlab. The phase signals are then further filtered with a 1MHz lowpass butterworth filter to approximate the final processing bandwidth. An FFT of the single side band (SSB) results are shown below in figure 6. The LO and Clock used in our system while fairly low noise should not be considered phase noise. By using the phase subtraction technique we get several orders of magnitude improvement by the subtraction of common noise. In fact, the LO and Clock can be run open loop with reduced performance. Further processing this data shows an RMS phase error of 0.0052 degrees integrated from 100 Hz to 1 MHz this was measured at an intermediate fan speed.

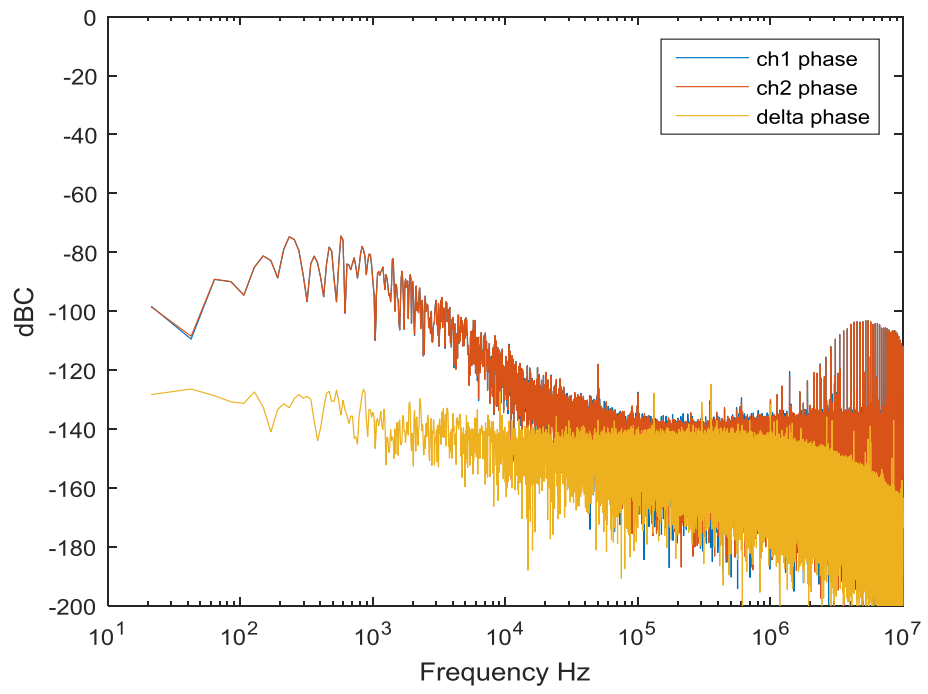

Figure 6: SSB Difference Phase Noise plot

**Amplitude Accuracy** Using the amplitude data from the algorithm mentioned in the phase accuracy section, we calculate at ch1 and ch2 amplitude errors at 0.0093 and -.0096%.

**Output Phase Noise** The output phase noise of the system must also meet very stringent requirements. To verify the performance we generated an output signal at 2856 MHz using prototype firmware. The SSB output phase noise as measured by an Agilent E5052B phase noise analyser is shown below in Fig. 7.

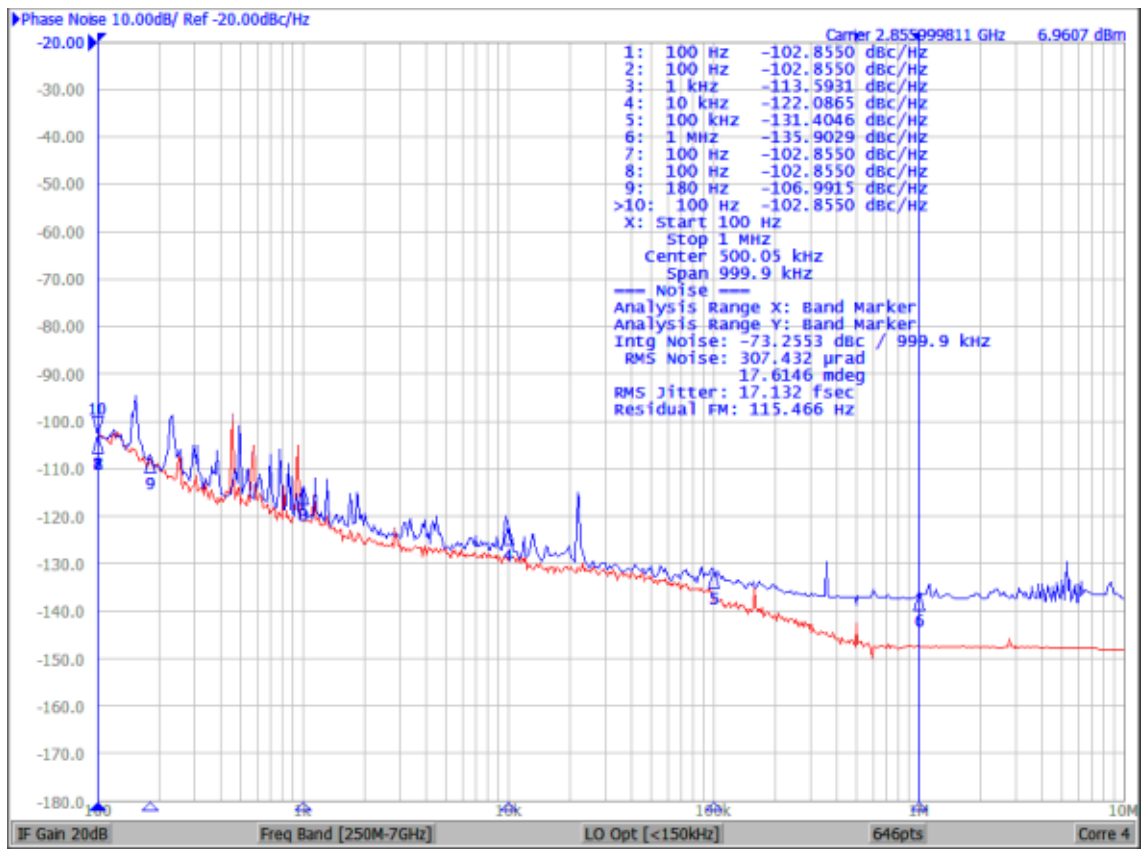

Figure 7: 2856 MHz output phase noise

The red line in Fig. 7, is the input reference noise. The blue line is the output phase noise which tracks the input phase noise quite well up to around 100 kHz at which point it becomes dominated by the noise floor of the output circuitry. This needs further investigation. The integrated rms noise from 100 Hz to 1 MHz is 0.0176 degrees.

**Other Performance Tests** One particular test meriting investigation is potential phase degradation under full ATCA fan speed. We ran the system at full fan speed and did, in fact, see a degradation of the phase noise. However, as can be seen in Fig 8, the full integrated phase noise is still below the 0.01 degree phase specification. To further mitigate this problem, we are planning on testing some dampening material on the oscillator daughter card as we expect that vibration on that card may be contributing to this added phase noise.

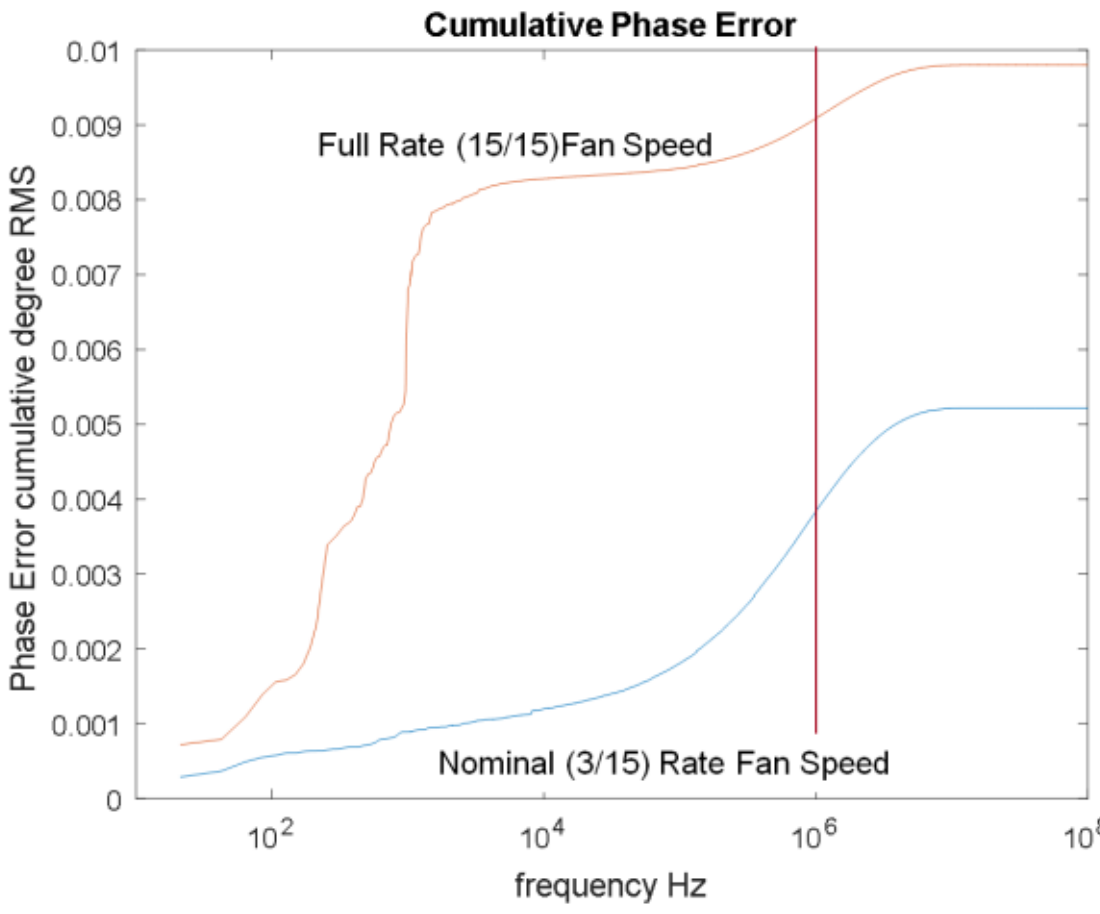

Figure 8: Phase Error at Full Fan Speed

## CONCLUSIONS

We have designed and tested prototype hardware for the ASU CXLS. Testing at SLAC LCLS-I shows that the system as currently built nearly meets all specifications.